\documentclass[a4]{article}

\usepackage{amsmath}
\usepackage{amsfonts}

\newcommand{\ol}{\overline}
\newcommand{\wt}{\widetilde}

\def\Det{\mathop{\rm Det}\nolimits}
\def\tr{\mathop{\rm tr}\nolimits}

\begin{document}
\begin{titlepage}
\title{
\begin{flushright}
\normalsize{ UT-11-01\\
TIT/HEP-607\\
Jan 2011}
\end{flushright}
       \vspace{2cm}
Index for three dimensional superconformal field theories
with general R-charge assignments
       \vspace{2cm}}
\author{
Yosuke Imamura\thanks{E-mail: \tt imamura@phys.titech.ac.jp}$^{~1}$
and Shuichi Yokoyama\thanks{E-mail: \tt yokoyama@hep-th.phys.s.u-tokyo.ac.jp}$^{~2}$
\\[30pt]
{\it $^1$ Department of Physics, Tokyo Institute of Technology,}\\
{\it Tokyo 152-8551, Japan}\\
{\it $^2$ Department of Physics, University of Tokyo,}\\
{\it Tokyo 113-0033, Japan}
}
\date{}

\maketitle
\thispagestyle{empty}

\vspace{0cm}

\begin{abstract}
\normalsize
We derive a general formula of an index
for three dimensional ${\cal N}=2$ superconformal field theories
with general R-charge assignments
to chiral multiplets
by using
the localization method in ${\bf S}^2\times{\bf S}^1$ background.
As examples we compute the index for theories in a few mirror pairs,
and confirm the agreement of the indices in each mirror pair.
\end{abstract}

\end{titlepage}


\section{Introduction}
In the last few years AdS$_4$/CFT$_3$ correspondence has attracted great interest.
Since the discovery of the ABJM model\cite{Aharony:2008ug},
which is dual to M-theory in AdS$_4\times{\bf S}^7/{\bf Z}_k$ and is
the first example of the duality,
many dual pairs with different numbers of supersymmetries
have been constructed.
Many non-trivial checks of the duality have been done.

The surprising prediction of AdS$_4$/CFT$_3$ duality,
the $N^{3/2}$ scaling\cite{Klebanov:1996un} of the partition function,
was recently confirmed on the CFT side
for theories with ${\cal N}\geq 3$ supersymmetry\cite{Drukker:2010nc,Herzog:2010hf}.
This is based on the fact that the path integral
of a three-dimensional ${\cal N}=2$ superconformal field theory in ${\bf S}^3$
can be reduced to a
matrix integral
by using the localization
method\cite{Kapustin:2009kz}.

In the derivation of the matrix model in \cite{Kapustin:2009kz},
chiral multiplets are assumed to have the canonical conformal dimension.
This is the reason why its applications had been
restricted until recent to ${\cal N}\geq 3$
theories, in which the R-symmetry is non-abelian and R-charges
of chiral multiplets are protected from quantum corrections.
However,
as is pointed out in \cite{Jafferis:2010un,Hama:2010av},
this is in fact not necessary for the localization procedure,
and actually it is possible to apply the localization
to theories including chiral multiplets with anomalous dimensions\cite{Jafferis:2010un,Hama:2010av}.
Furthermore,
it is proposed in \cite{Jafferis:2010un} that we can determine R-charges
of chiral multiplets at infra-red fixed points by
extremizing the ${\bf S}^3$ partition function.
This is an important development because this enables us to
compute the partition functions
of a very large class of theories
and to confirm predictions of AdS$_4$/CFT$_3$ for many interesting examples.

There exists another quantity which can be computed exactly.
It is the ${\cal N}=2$ superconformal index 
\begin{equation}
I(x_2,y_a,z_i)=\tr\left[
(-1)^Fx_1^{\Delta-R-j_3}x_2^{\Delta+j_3}y_a^{m_a}z_i^{F_i}
\right],
\label{indexedf}
\end{equation}
for three-dimensional theories compactified on ${\bf S}^2$.
$\Delta$, $R$, $j_3$ and $F_i$ are
the energy, the R-charge, the third component of the angular momentum
rotating ${\bf S}^2$,
and flavor charges, respectively.
$m_a$ are magnetic charges associated with $U(1)$ gauge groups labeled by the index $a$.
This index does not depend on the variable $x_1$.
The reason is as follows.
The operator $\Delta-R-j_3$ in the exponent of $x_1$
in (\ref{indexedf}) is written in the form
\begin{equation}
\Delta-R-j_3=\{{\cal Q}^\dagger,{\cal Q}\}
\label{qqdagger}
\end{equation}
where ${\cal Q}$ is one of eight supercharges
in the ${\cal N}=2$ superconformal algebra.
It is easily shown that
only ${\cal Q}$ and ${\cal Q}^\dagger$-invariant BPS states contribute to the
index, and thus the index does not depend on $x_1$.
We can exactly compute this index by using
the localization technique again.
The index contains different information from
the ${\bf S}^3$ partition function, and
it would be important to use them complementarily.
In particular, the index is useful to study the field-operator correspondence.
For such studies for AdS$_4$/CFT$_3$ with ${\cal N}\geq3$ supersymmetry, see \cite{Bhattacharya:2008zy,Bhattacharya:2008bja,Choi:2008za,Kim:2009wb,Imamura:2009hc,Kim:2010vwa}.

Until now the index is only computed for theories
without anomalous dimensions.
However, just like the partition function,
it is in fact possible to compute the index for theories with
anomalous dimensions
if we know the R-charge assignments.
The purpose of this paper is to derive a
general formula for the index in such a case.
We follow the prescription given in \cite{Kim:2009wb},
and extend the result in \cite{Kim:2009wb} to theories
with chiral multiplets with non-canonical R-charges.
We will give details only for chiral multiplets.
The reader should refer to \cite{Kim:2009wb} for details
about the contribution of vector multiplets.
See also \cite{Gang:2009wy} for partition functions
of vector multiplets on general lens spaces $L(p,q)$ including
${\bf S}^3$ and ${\bf S}^2\times{\bf S}^1$ as special cases.

This paper is organized as follows.
In the next section we review
how we obtain deformation terms
in the action
necessary for the localization.
In section \ref{s2s1} we consider
theories in ${\bf S}^2\times {\bf S}^1$
and explicitly give the
deformation terms for the background.
We carry out the Gaussian integral in section \ref{gaussian}
and obtain a general formula for the index.
We apply the formula to some examples in section \ref{examples}.
Section \ref{discussion} is devoted to discussions.

\section{Supersymmetry in conformally flat backgrounds}
We consider an ${\cal N}=2$ superconformal field theory in a conformally
flat Euclidean background
with gauge group $G$ and an arbitrary number of chiral multiplets.
The partition function is defined by the path integral
\begin{equation}
Z=\int{\cal D}\Psi e^{-S[\Psi]},
\label{zdef}
\end{equation}
where $S[\Psi]$ is the Euclidean action of
the theory and
$\Psi$ denotes arbitrary fields in the theory.
An appropriate gauge fixing in 
(\ref{zdef}) is understood.
To compute the path integral in
(\ref{zdef}) exactly by using the localization
technique,
we choose one supersymmetry $\delta_1$ respected by the background,
and deform the action by adding $\delta_1$-exact terms $S_{\rm def}=t\delta_1 V$
($t\in{\bf R}$).
$Z$ does not depend on
the deformation parameter $t$.
If we take an appropriate
$S_{\rm def}$, the path integral of all but finite degrees of freedom
reduces to Gaussian integral in the $t\rightarrow \infty$ limit.

The ${\cal N}=2$ superconformal algebra contains
$8$ supercharges; four of them are parameterized by a complex spinor $\epsilon$,
and the other four by another complex spinor $\ol\epsilon$.
In a Euclidean space these spinors should be treated
as independent ones.
In a conformally flat background
the spinor $\epsilon$ satisfies
the Killing equation\cite{Kapustin:2009kz}
\begin{equation}
D_\mu\epsilon=\gamma_\mu\kappa,
\label{killingeq}
\end{equation}
with an arbitrary spinor $\kappa$.
In the flat background,
general solution to (\ref{killingeq}) is
$\epsilon=\xi-x^\mu\gamma_\mu\zeta$ with constant spinors $\xi$ and $\zeta$
corresponding to $Q$ and $S$ transformations, respectively.
By using $\epsilon$, $Q$ and $S$ transformations are
given in a unified way.
This is also the case for the anti-holomorphic parameter $\ol\epsilon$,
corresponding to $\ol Q$ and $\ol S$.
For the computation of the ${\bf S}^3$ partition function,
we can use any of holomorphic or antiholomorphic supersymmetries
as $\delta_1$,
whereas we should use an anti-holomorphic supersymmetry
when we compute the index encoding the spectrum of BPS operators.
The supercharge ${\cal Q}$ in (\ref{qqdagger}) should be
a component of $\ol Q$ or $\ol S$.

For a vector multiplet $(A_\mu,\sigma,D,\lambda)$
the anti-holomorphic part of transformation laws are
\begin{eqnarray}
&&
\delta\sigma=(\ol\epsilon\lambda),\quad
\delta A_\mu=-i(\ol\epsilon\gamma_\mu\lambda),\quad
\delta\lambda=0,
\nonumber\\&&
\delta D=
i(\ol\epsilon\gamma^\mu D_\mu\lambda)
+i(\ol\epsilon[\sigma,\lambda])
+\frac{i}{3}(D_\mu\ol\epsilon\gamma^\mu\lambda),
\nonumber\\&&
\delta\ol\lambda=-\frac{i}{2}\gamma^{\mu\nu}\ol\epsilon F_{\mu\nu}
-\gamma^\mu\ol\epsilon D_\mu\sigma
+iD\ol\epsilon
-\frac{2}{3}\gamma^\mu D_\mu\ol\epsilon \sigma.
\label{sstrqs}
\end{eqnarray}
The transformation laws for a
chiral multiplet $\Phi=(\phi,\psi,F)$ with Weyl weight $\Delta_\Phi$
are
\begin{eqnarray}
&&
\delta\phi^\dagger=\sqrt2(\ol\epsilon\ol\psi),\quad
\delta\phi=0,\quad
\delta\ol\psi=\sqrt2i\ol\epsilon F^\dagger,\quad
\delta F^\dagger=0,
\nonumber\\&&
\delta\psi=
\sqrt2\ol\epsilon \sigma\phi
-\sqrt2\gamma^\mu\ol\epsilon D_\mu\phi
-\frac{2\sqrt2}{3}\Delta_\Phi\phi\gamma^\mu D_\mu\ol\epsilon,
\nonumber\\&&
\delta F=
\sqrt2i(\ol\epsilon\gamma^\mu D_\mu\psi)
+\sqrt2i(\ol\epsilon\sigma\psi)
+2i(\ol\epsilon\ol\lambda)\phi
+\frac{2\sqrt2i}{3}\left(\Delta_\Phi-\frac{1}{2}\right)
     (D_\mu\ol\epsilon\gamma^\mu\psi).
\label{chiraltr}
\end{eqnarray}
The R-charge of a chiral multiplet $\Phi$ is the same as the Weyl weight
$\Delta_\Phi$.
The transformation laws (\ref{sstrqs}) and (\ref{chiraltr})
can be used not only for the flat background
but also for general conformally flat backgrounds,
including ${\bf S}^3$ and ${\bf S}^2\times{\bf S}^1$.

As we mentioned above,
for the computation of the path integral,
we need to deform the action by
$\delta_1$-exact terms.
Let us first consider flat ${\bf R}^3$.
In this case we choose one particular
constant spinor $\ol\epsilon_1$ to define the supersymmetry $\delta_1$,
and we use
\begin{eqnarray}
&&S^{{\rm vector},{\bf R}^3}_{\rm def}
=\int d^3x\int d^2\ol\theta \tr\left(-\frac{t}{2}\ol W\ol W\right)
\nonumber\\
&&=
t\int d^3x
\tr\left[\frac{1}{2}F_{\mu\nu}F^{\mu\nu}
+D_\mu\sigma  D^\mu\sigma
+D^2
-2(\ol\lambda\gamma^\mu D_\mu\lambda)
-2(\ol\lambda[\sigma,\lambda])\right].
\label{s0v}
\end{eqnarray}
for the deformation of the action
of a vector multiplet.
The trace ``tr'' should be a positive definite $G$-invariant inner product.
The action $S^{{\rm vector},{\bf R}^3}_{\rm def}$ given above is $\delta_1$-exact
by construction.
Indeed, we can rewrite (\ref{s0v}) as
\begin{equation}
S^{\rm vector}_{\rm def}
=\delta_1\delta_2\int \sqrt{g}\tr\left(-\frac{t}{2}\ol\lambda\ol\lambda\right)d^3x,
\label{intth22}
\end{equation}
where $\delta_2$ is a supersymmetry defined with
another constant spinor $\ol\epsilon_2$
linearly independent of $\ol\epsilon_1$.
We insert the volume factor $\sqrt{g}$
because we will use (\ref{intth22}) for ${\bf S}^3$ and
${\bf S}^2\times{\bf S}^1$, too.

For a chiral multiplet, we deform the theory by
\begin{eqnarray}
&&S^{{\rm chiral},{\bf R}^3}_{\rm def}
=\int d^3x\int d^2\ol\theta D^2\left(-\frac{t}{2}\Phi^\dagger e^V\Phi\right)
\nonumber\\
&&
=t\int d^3x\Big[-\phi^\dagger D_\mu D^\mu\phi
-(\ol\psi\gamma^\mu D_\mu\psi)  
+F^\dagger F
+\phi^\dagger\sigma\sigma\phi
+i\phi^\dagger D\phi
\nonumber\\&&\hspace{4cm}
-\sqrt2(\ol\psi\ol\lambda)\phi
-\sqrt2\phi^\dagger(\lambda\psi)
-(\ol\psi\sigma\psi)\Big].
\label{s0c}
\end{eqnarray}
This is again rewritten with $\delta_1$ and $\delta_2$ as
\begin{equation}
S^{\rm chiral}_{\rm def}
=\delta_1\delta_2\int\sqrt{g}\left(-\frac{it}{2}\phi^\dagger F\right) d^3x.
\label{ddfp}
\end{equation}

(\ref{intth22}) and (\ref{ddfp}) can be used to construct
deformation terms for theories
in ${\bf S}^3$ and ${\bf S}^2\times{\bf S}^1$, too.
In the case of ${\bf S}^3$ with radius $r$,
four independent anti-holomorphic killing spinors
split into two satisfying $D_\mu\ol\epsilon=(i/2r)\gamma_\mu\ol\epsilon$
and the other two satisfying $D_\mu\ol\epsilon=-(i/2r)\gamma_\mu\ol\epsilon$.
For the construction of deformation terms,
we use the former two or the latter two.
Let us choose the former two satisfying
\begin{equation}
D_\mu\ol\epsilon=\frac{i}{2r}\gamma_\mu\ol\epsilon.
\end{equation}
We denote two linearly independent killing spinors satisfying this equation
by $\ol\epsilon_1$ and $\ol\epsilon_2$,
and the corresponding supersymmetries by $\delta_1$ and $\delta_2$.
$\delta_1$-exact deformation actions in ${\bf S}^3$ are given by
(\ref{intth22}) and (\ref{ddfp}), again.
Due to the modification in the transformation laws,
additional terms including $1/r$ arise in the action.
\begin{eqnarray}
&&S^{{\rm vector},{\bf S}^3}_{\rm def}
=
S^{{\rm vector},{\bf R}^3}_{\rm def}
+t\int d^3x\sqrt{g}\left[
-\frac{2}{r}D\sigma
+\frac{1}{r^2}\sigma^2
-\frac{i}{r}(\ol\lambda\lambda)\right],\nonumber\\
&&S^{{\rm chiral},{\bf S}^3}_{\rm def}
=
S^{{\rm chiral},{\bf R}^3}_{\rm def}
\nonumber\\&&\hspace{3em}
+t\int d^3x\sqrt{g}\left[
\frac{i(2\Delta_\Phi-1)}{2r}(\ol\psi\psi-2\phi^\dagger\sigma\phi)
-\frac{\Delta_\Phi(\Delta_\Phi-2)}{r^2}\phi^\dagger\phi
\right].
\label{svsv}
\end{eqnarray}
$S^{{\rm vector},{\bf R}^3}_{\rm def}$ and
$S^{{\rm chiral},{\bf R}^3}_{\rm def}$
in (\ref{svsv}) are
the actions (\ref{s0v}) and (\ref{s0c})
with appropriate covariantization of the integration measure
and the derivatives.
By using these action, we can compute the exact partition function
in ${\bf S}^3$\cite{Kapustin:2009kz,Jafferis:2010un,Hama:2010av}.

\section{Localization in ${\bf S}^2\times {\bf S}^1$}\label{s2s1}
The purpose of this paper is to compute the index (\ref{indexedf})
as the path integral (\ref{zdef}) of a theory
defined on the background ${\bf S}^2\times{\bf S}^1$.
Let $r$ and $\beta r$ be the radius of ${\bf S}^2$ and the
period of ${\bf S}^1$, respectively.
We use coordinates $x^i$ ($i=1,2$) for ${\bf S}^2$ and $x^3$ for ${\bf S}^1$.
The deformation terms necessary for the localization
are given by (\ref{intth22}) and (\ref{ddfp}) with
an appropriate choice of Killing spinors $\ol\epsilon_1$ and $\ol\epsilon_2$.
Before considering the compact space ${\bf S}^2\times{\bf S}^1$,
let us consider Killing spinors in ${\bf S}^2\times{\bf R}$.
Four linearly independent Killing spinors in this non-compact background
split into two satisfying
\begin{equation}
D_\mu\ol\epsilon=\frac{1}{2r}\gamma_\mu\gamma_3\ol\epsilon
\label{s2s1killing}
\end{equation}
and the other two satisfying the equation with opposite sign.
Let $\ol\epsilon_1$ and $\ol\epsilon_2$ be
two linearly independent spinors
satisfying (\ref{s2s1killing}), and
$\delta_1$ and $\delta_2$ be the corresponding supersymmetry transformations.
These two spinors form a doublet of $SO(3)$ rotation symmetry
of ${\bf S}^2$.
We define the $SO(3)$ angular momentum $j$ ($j\geq0$) and
its third component $j_3$ ($-j\leq j_3\leq j$).
We assume
$\ol\epsilon_1$ and $\ol\epsilon_2$ have $j_3$ eigenvalues
$+1/2$ and $-1/2$, respectively.

Let us now consider how we can compactify the ``time'' direction
${\bf R}$ to ${\bf S}^1$.
The Killing equation (\ref{s2s1killing})
implies $\ol\epsilon_1\propto e^{x^3/(2r)}$,
and we cannot impose the periodic boundary condition
on $\ol\epsilon_1$.
Instead, $\ol\epsilon_1$ satisfies
\begin{equation}
\ol\epsilon_1(x^3+\beta r)=e^{\beta/2}\ol\epsilon_1(x^3).
\label{epsilonbc0}
\end{equation}
We interpret the extra factor $e^{\beta/2}$ on the right hand side
as an insertion of a twist operator.
Namely,
by using the quantum numbers
\begin{equation}
R(\ol\epsilon_1)=-1,\quad
j_3(\ol\epsilon_1)=\frac{1}{2},\quad
F_i(\ol\epsilon_1)=0,
\end{equation}
we can rewrite (\ref{epsilonbc0}) as
\begin{equation}
\ol\epsilon_1(x^3+\beta r)
=e^{(-R-j_3)\beta_1+j_3\beta_2+F_i\gamma_i}\ol\epsilon_1(x^3),
\end{equation}
where
$\beta_1$, $\beta_2$, and $\gamma_i$ are real parameters
satisfying $\beta=\beta_1+\beta_2$.
For the consistency, the same boundary condition should be imposed
on all fields in the theory.
Namely, we impose
\begin{equation}
\Psi(x^3+\beta r)=
e^{(-R-j_3)\beta_1+j_3\beta_2+F_i\gamma_i}
\Psi(x^3)
\label{s1bc}
\end{equation}
for an arbitrary field $\Psi$.
We can also insert the factor $y_a^{m_a}$ when there exist $U(1)$ gauge groups,
but it does not change the boundary conditions for elementary fields.
The path integral over ${\bf S}^2\times{\bf S}^1$ with this twisted boundary condition
gives the index $I(x_2,y_a,z_i)$ defined in (\ref{indexedf}).
The variables $x_1$, $x_2$, and $z_i$ are
related to $\beta_1$, $\beta_2$, and $\gamma_i$
by
\begin{equation}
x_1=e^{-\beta_1},\quad
x_2=e^{-\beta_2},\quad
z_i=e^{-\gamma_i}.
\label{variablexxz}
\end{equation}

Deformation actions given by
(\ref{intth22}) and (\ref{ddfp})
with the killing spinors $\ol\epsilon_1$ and $\ol\epsilon_2$
satisfying (\ref{s2s1killing}) are
\begin{equation}
S^{{\rm vector},{\bf S}^2\times{\bf S}^1}_{\rm def}
=t\int d^3x\sqrt{g}\tr\left[
V_\mu V^\mu+D^2
-2(\ol\lambda\gamma^\mu D_\mu\lambda)
-2(\ol\lambda[\sigma,\lambda])
-\frac{1}{r}(\ol\lambda\gamma_3\lambda)
\right],
\label{vectordef}
\end{equation}
for a vector multiplet and
\begin{eqnarray}
&&\hspace{-1cm}
S^{{\rm chiral},{\bf S}^2\times{\bf S}^1}_{\rm def}
=S^{{\rm chiral},{\bf R}^3}_{\rm def}
\nonumber\\&&
+t\int d^3x\sqrt{g}\left[
\frac{1-2\Delta_\Phi}{r}
\left(\phi^\dagger D_3\phi
+\frac{1}{2}(\ol\psi\gamma_3\psi)\right)
+\frac{\Delta_\Phi(1-\Delta_\Phi)}{r^2}\phi^\dagger\phi
\right],
\label{chiraldef}
\end{eqnarray}
for a chiral multiplet with Weyl weight $\Delta_\Phi$.
The vector $V_\mu$ in
(\ref{vectordef}) is defined by
\begin{equation}
V_1=F_{23}-D_1\sigma,\quad
V_2=F_{31}-D_2\sigma,\quad
V_3=F_{12}-D_3\sigma-\frac{1}{r}\sigma.
\end{equation}

The path integral localizes around
$V_\mu=0$.
The equations $V_\mu=0$ are solved by
\begin{equation}
A_\mu^{(0)} dx^\mu=\frac{a}{\beta r}dx^3+mB_idx^i,\quad
\sigma^{(0)}=\frac{m}{2r}=\frac{s}{r}.
\label{a0s0}
\end{equation}
Expectation values of all other fields vanish.
We assume that the Gaussian integral is
dominated by GNO monopoles.
$B_i$ is the Dirac monopole configuration with unit magnetic charge.
After an appropriate gauge fixing,
$a$, $m$, and $s$ take values in the Cartan part of the Lie algebra
of the gauge group $G$.
$a$ is the Wilson line around ${\bf S}^1$, and
$m$ is the magnetic charge of the monopole.
$s=m/2$ is defined for later convenience.
When we perform the Gaussian integral,
all fields in this background is decomposed
into vacuum expectation values $\Psi^{(0)}$ and the fluctuations $\Psi'$
as
\begin{equation}
\Psi=\Psi^{(0)}+\frac{1}{\sqrt{t}}\Psi'.
\label{psi0psip}
\end{equation}
Substituting this into the total action $S+S_{\rm def}$,
all interaction terms including more than two fluctuations
vanish in $t\rightarrow\infty$ limit.
After taking the limit,
we are left with
\begin{equation}
I=\sum_{m}y_a^{m_a}\int da\int {\cal D}\Psi' e^{-S^{(0)}}e^{-\int\sqrt{g}\Psi'D\Psi'd^3x},
\end{equation}
where $D$ is a certain differential operator acting on the fluctuations
and $S^{(0)}$ is the expectation value of the original action.
Almost all terms in the original action vanish when the
expectation values are substituted.
If the action contains Chern-Simons terms, it gives
the non-vanishing contribution
\begin{equation}
S_{\rm CS}^{(0)}
=\frac{i}{4\pi}\int\tr'\left(A^{(0)}dA^{(0)}-\frac{2i}{3}A^{(0)}A^{(0)}A^{(0)}
\right)
=2i\tr'(as).
\end{equation}
The definition of the trace ``$\tr'$'' here includes Chern-Simons
levels and it does not have to be positive definite.

In order to perform the Gaussian integral
with respect to the fluctuations $\Psi'$,
we expand the fields by spherical harmonics on ${\bf S}^2$.
Let us focus on a particular component of $\Psi'$
belonging to a representation $R_\Phi$ of $G$.
Its spin on ${\bf S}^2$ and its weight in the representation $R_\Phi$
are denoted by $S$ and $\rho$, respectively.
The covariant derivative on ${\bf S}^2$ acting on the component
is given by
\begin{equation}
D_i=\partial_i-i\rho(m)B_i-iS\omega_i,
\end{equation}
where $\omega_i$ is the spin connection on ${\bf S}^2$.
Because the spin connection $\omega$ and the monopole potential
$B$ on ${\bf S}^2$ satisfy the relation
\begin{equation}
B=\frac{1}{2}\omega,
\end{equation}
we can rewrite the covariant derivative as
\begin{equation}
D_i=\partial_i-iS_{\rm eff}\omega_i,
\end{equation}
with the effective spin
\begin{equation}
S_{\rm eff}
=S+\frac{1}{2}\rho(m)
=S+\rho(s).
\end{equation}
Therefore, we can use spin $S_{\rm eff}$ spherical harmonics
\begin{equation}
Y^{S_{\rm eff}}_{j,j_3},\quad
j\geq |S_{\rm eff}|,\quad
-j\leq j_3\leq j,
\end{equation}
for the expansion of the component with
spin $S$ and weight $\rho$ in the background flux $m$.
This fact is quite useful to compute the eigenvalues of
differential operators.

\section{Gaussian integral}\label{gaussian}
Let us carry out the Gaussian integral by using the deformed action.
We consider a chiral multiplet $\Phi$
with Weyl weight $\Delta_\Phi$ belonging to a representation $R_\Phi$ of $G$.
After substituting (\ref{psi0psip}) and
taking the limit $t\rightarrow\infty$,
the action becomes
\begin{eqnarray}
S^{\rm chiral,{\bf S}^2\times{\bf S}^1}_{\rm def}&=&
-\phi^\dagger D_\mu D^\mu\phi
+\frac{1}{r^2}\phi^\dagger ss\phi
+\frac{\Delta_\Phi(1-\Delta_\Phi)}{r^2}\phi^\dagger\phi
+\frac{1-2\Delta_\Phi}{r}\phi^\dagger D_3\phi
\nonumber\\&&
-\frac{1}{r}(\ol\psi s\psi)
-(\ol\psi\gamma^\mu D_\mu\psi)  
+\frac{1-2\Delta_\Phi}{2r}(\ol\psi\gamma_3\psi)
\nonumber\\&&
+F^\dagger F,
\label{freeparts}
\end{eqnarray}
where the gauge fields in the covariant derivatives
are replaced by $A_\mu^{(0)}$.
All fields in (\ref{freeparts}) represent
the fluctuation part $\Psi'$ in (\ref{psi0psip}).
The integration of the auxiliary field $F$ gives
a constant factor and we can simply drop it.

The path integral of the complex scalar field $\phi$ gives
the factor $Z_\phi=(\Det D_\phi)^{-1}$ with
the differential operator
\begin{equation}
D_\phi
=
-D_3 D_3
-D_i D_i
+\frac{1}{r^2}s^2
+\frac{\Delta_\Phi(1-\Delta_\Phi)}{r^2}
+\frac{1-2\Delta_\Phi}{r}D_3.
\label{dphi}
\end{equation}
Let us focus on a component
of the scalar field with weight $\rho\in R_\Phi$.
Although the spin of scalar field is $S=0$,
the coupling to the background flux shifts the effective
spin to $S_{\rm eff}=\rho(s)$.
We can expand such a field by spin $S_{\rm eff}$ spherical
harmonics $Y^{S_{\rm eff}}_{j,j_3}$.
The eigenvalue of the Laplacian $D_iD_i$ corresponding to
$Y^{S_{\rm eff}}_{j,j_3}$ is
\begin{equation}
D_iD_iY^{S_{\rm eff}}_{j,j_3}
=-\frac{1}{r^2}[j(j+1)-S_{\rm eff}^2]Y^{S_{\rm eff}}_{j,j_3}.
\label{laplacian}
\end{equation}
Substituting (\ref{laplacian}) into
(\ref{dphi}), we obtain the eigenvalue
\begin{equation}
D_\phi
=\frac{1}{r^2}
(j+\Delta_\Phi+rD_3)(j+1-\Delta_\Phi-rD_3).
\end{equation}
In this expression $D_3$ should be understood to be its
eigenvalue.
By taking the twisted boundary condition
(\ref{s1bc}) into account,
the eigenvalues of $D_3$ are given by
\begin{equation}
D_3=
\frac{1}{\beta r}
\Big[
2\pi in-i\rho(a)+(-R-j_3)\beta_1+j_3\beta_2+F_i\gamma_i
\Big],\quad
n\in {\bf Z}.
\label{d3eigen}
\end{equation}
For the scalar field $\phi$ the R-charge $R$ in (\ref{d3eigen}) is replaced
by $\Delta_\Phi$.
Taking the product of all the eigenvalues, we obtain
the scalar field contribution to the Gaussian integral,
\begin{equation}
Z_\phi
=\left[
\prod_{\rho\in R_\Phi}
\prod_{j=|\rho(s)|}^\infty\prod_{j_3=-j}^j
\prod_{n=-\infty}^\infty(j+\Delta_\Phi+rD_3)(j+1-\Delta_\Phi-rD_3)
\right]^{-1}.
\end{equation}

Next, let us consider Gaussian integral of the fermion field $\psi$.
The differential operator acting on $\psi$ in the action
(\ref{freeparts}) is
\begin{equation}
D_{\rm fer}=\gamma^\mu D_\mu-\frac{1-2\Delta_\Phi}{2r}\gamma_3+\frac{s}{r}.
\end{equation}
We focus on a component with weight $\rho\in R_\Phi$.
Including the shift due to the background flux,
the upper and lower components of the spinor
$\psi$ have the effective spins
$S_{\rm eff}=\rho(s)-1/2$ and
$S_{\rm eff}=\rho(s)+1/2$, respectively.
They are expanded by spherical harmonics
$Y^{\rho(s)-1/2}_{j,j_3}$
and $Y^{\rho(s)+1/2}_{j,j_3}$.
Let us focus on a mode with spin $j$.

When $j\geq|\rho(s)|+1/2$, both
$Y^{\rho(s)-1/2}_{j,j_3}$ and $Y^{\rho(s)+1/2}_{j,j_3}$ exist,
and the differential operator $D_{\rm fer}$ acting on $\psi$
takes the matrix form
\begin{equation}
D_{\rm fer}
=\left(\begin{array}{cc}
D_3-\frac{1-2\Delta_\Phi}{2r}+\frac{\rho(s)}{r} & D_+ \\
D_-        & -D_3+\frac{1-2\Delta_\Phi}{2r}+\frac{\rho(s)}{r}
\end{array}\right),
\label{dferc}
\end{equation}
where $D_\pm=D_1\pm iD_2$.
The determinant of the matrix (\ref{dferc}) is
\begin{equation}
\det D_{\rm fer}=
\frac{\rho(s)^2}{r^2}-\left(D_3-\frac{1-2\Delta_\Phi}{2r}\right)^2
-D_+D_-.
\label{detdiffop}
\end{equation}
``$\det$'' in (\ref{detdiffop}) represents the determinant
of the $2\times 2$ matrix,
while
``$\Det$'' is used for the determinant of differential operators.
Note that $D_+$ and $D_-$ do not commute with each other
and $D_+D_-$ and $D_-D_+$ are different operators.
If we adopt $D_+D_-$ as in (\ref{detdiffop})
we should regard it as an operator acting on
the upper component of $\psi$,
which has the effective spin $\rho(s)-1/2$.
The eigenvalue is
\begin{equation}
D_+D_-Y^{\rho(s)-1/2}_{j,j_3}
=-\frac{1}{r^2}\left[\left(j+\frac{1}{2}\right)^2-\rho(s)^2\right]Y^{\rho(s)-1/2}_{j,j_3}.
\label{ddeigen}
\end{equation}
We can also use $D_-D_+$
acting on $Y^{\rho(s)+1/2}_{j,j_3}$,
and obtain the same eigenvalue as (\ref{ddeigen}).
By substituting this eigenvalue into 
(\ref{detdiffop}) we obtain
\begin{equation}
\det D_{\rm fer}
=\frac{1}{r^2}(j+\Delta_\Phi+rD_3)(j+1-\Delta_\Phi-rD_3).
\label{dfee}
\end{equation}

If $j=|\rho(s)|-1/2$,
only one of $Y^{\rho(s)-1/2}_{j,j_3}$ or $Y^{\rho(s)+1/2}_{j,j_3}$ exists,
and thus only top-left or bottom-right component
in the matrix (\ref{dferc}) exists.
The eigenvalue in this case is
\begin{equation}
D_{\rm fer}=\frac{1}{r}(j+\Delta_\Phi+rD_3).
\label{dfe}
\end{equation}

Combining (\ref{dfee}) and (\ref{dfe}),
we obtain
\begin{eqnarray}
Z_\psi=\Det D_{\rm fer}
&=&
\prod_{\rho\in R_\Phi}\prod_{j=|\rho(s)|-1/2}^\infty
\prod_{j_3=-j}^j\prod_{n=-\infty}^\infty
(j+\Delta_\Phi+rD_3)
\nonumber\\&\times&
\prod_{\rho\in R_\Phi}\prod_{j=|\rho(s)|+1/2}^\infty
\prod_{j_3=-j}^j\prod_{n=-\infty}^\infty
(j+1-\Delta_\Phi-rD_3)
\end{eqnarray}
Recall that the differential operator $D_3$
should be understood as its eigenvalue
given in (\ref{d3eigen}).
For the fermion field $\psi$
the R-charge $R$ in (\ref{d3eigen})
is replaced by $\Delta_\Phi-1$.

A similar contribution is also obtained from
vector multiplets.
Let us denote it by $Z_{\rm vector}$.
See \cite{Kim:2009wb} for its explicit form and a
detailed derivation.

We now obtain the following expression for the index
\begin{equation}
I=\sum_sy_a^{2s_a}\int da
e^{-S^{(0)}_{\rm CS}}
Z_{\rm vector}
\prod_\Phi(Z_\phi Z_\psi),
\label{izzz}
\end{equation}
$\prod_\Phi$ represents the product over all chiral multiplets.
$Z$'s in the integrand in (\ref{izzz})
contain infinite product of eigenvalues.
These are treated in a standard way.
Let us focus on the factor
$(j+\Delta_\Phi+rD_3)$ in $Z_\phi$.
The explicit form of this eigenvalue is
\begin{equation}
\beta (j+\Delta_\Phi+rD_3)
=2\pi in-i\rho(a)+(j-j_3)\beta_1+(j+\Delta_\Phi+j_3)\beta_2+F_i\gamma_i
\label{jdp}
\end{equation}
Let $z$ be the right hand side with $2\pi in$ removed.
We also define $z$ in the same way for other series of eigenvalues in
(\ref{izzz}).
Then the product of eigenvalues
in (\ref{izzz}) can be written as
\begin{equation}
\prod_{\cdots}\prod_{n=-\infty}^\infty(2\pi in+z)^{-(-)^F},
\end{equation}
where the first product $\prod_{\cdots}$ represents all the products but one
with respect to $n$.
$F$ is the fermion number of the corresponding field.

We first carry out the product over the integer $n$
by using the formula
\begin{eqnarray}
\prod_{n=-\infty}^\infty(2\pi i n+z)
=
2\sinh\frac{z}{2}
=e^\frac{z}{2}(1-e^{-z})
=e^\frac{z}{2}\exp\left[
-\sum_{m=1}^\infty\frac{1}{m}e^{-mz}
\right].
\end{eqnarray}
At the first equality we neglect a divergent constant.
With this formula
the product $\prod_{\cdots}$
in the definition of
$Z$
can be rewritten by the summation $\sum_{\cdots}$,
\begin{eqnarray}
\prod_{\cdots}\prod_{n=-\infty}^\infty(2\pi i n+z)^{-(-)^F}
=e^{-\sum_{\cdots}(-)^F\frac{z}{2}}\exp\left[
\sum_{m=1}^\infty\frac{1}{m}\sum_{\cdots}(-)^F e^{-mz}
\right].
\label{pplambda}
\end{eqnarray}
We define th function
\begin{equation}
f(e^{ia},x_1,x_2,z_i)=\sum_{\cdots}(-)^F e^{-z}.
\end{equation}
We call this the letter index because this
can be regarded as an index for elementary excitations,
which are often called letters.
For the eigenvalue (\ref{jdp}), $e^{-z}$ is given by
\begin{equation}
e^{-z}
=e^{i\rho(a)-(j-j_3)\beta_1-(j+\Delta_\Phi+j_3)\beta_2-F_i\gamma_i}
=e^{i\rho(a)}x_1^{j-j_3}x_2^{j+\Delta_\Phi+j_3}z_i^{F_i},
\end{equation}
and the corresponding letter index is
\begin{equation}
f=\sum_{\rho,j,j_3}e^{-z}
=\sum_{\rho\in R_\Phi}e^{i\rho(a)}x_2^{\Delta_\Phi} z_i^{F_i}
\sum_{j=|\rho(s)|}^\infty
\sum_{j=-j_3}^{j_3}
(x_1x_2)^j\left(\frac{x_2}{x_1}\right)^{j_3}.
\label{fc1}
\end{equation}

We need to compute the letter index
for other series of eigenvalues in the integrand
in (\ref{izzz}) in the same way.
We give only the results.
From the other factor $(j+1-\Delta_\Phi-rD_3)$ in $Z_\phi$
we obtain
\begin{equation}
f=
\sum_{\rho\in R_\Phi}e^{-i\rho(a)}x_2^{-\Delta_\Phi}z_i^{-F_i}
\sum_{j=|\rho(s)|}^\infty\sum_{j_3=-j}^j
(x_1x_2)^{j+1}\left(\frac{x_1}{x_2}\right)^{j_3}.
\label{fc2}
\end{equation}
The factor $(j+\Delta_\Phi+rD_3)$ in $Z_\psi$ gives
\begin{equation}
f
=-\sum_{\rho\in R_\Phi}e^{i\rho(a)}x_2^{\Delta_\Phi} z_i^{F_i}\sum_{k=|\rho(s)|}^\infty\sum_{l=-k}^{k-1}
(x_1x_2)^k\left(\frac{x_2}{x_1}\right)^l,
\label{fc3}
\end{equation}
and the other factor $(j+1-\Delta_\Phi-rD_3)$ in $Z_\psi$ gives
\begin{equation}
f
=-\sum_{\rho\in R_\Phi}e^{-i\rho(a)}x_2^{-\Delta_\Phi}z_i^{-F_i}\sum_{k=|\rho(s)|}^\infty\sum_{l=-k-1}^{k}
(x_1x_2)^{k+1}\left(\frac{x_1}{x_2}\right)^l.
\label{fc4}
\end{equation}
By summing up (\ref{fc1}), (\ref{fc2}), (\ref{fc3}), and (\ref{fc4})
we obtain the letter index for a chiral multiplet $\Phi$.
Finally we sum up the contributions of all chiral multiplets,
and obtain
\begin{equation}
f_{\rm chiral}(e^{ia},x_2,z_i)=
\sum_\Phi\sum_{\rho\in R_\Phi}
\left[
e^{i\rho(a)}z_i^{F_i}\frac{x_2^{2|\rho(s)|+\Delta_\Phi}}{1-x_2^2}
-e^{-i\rho(a)}z_i^{-F_i}
\frac{x_2^{2|\rho(s)|+2-\Delta_\Phi}}{1-x_2^2}
\right],
\label{fchiral}
\end{equation}
where $\sum_\Phi$ represents the summation over all chiral multiplets.
This does not depend on the variable $x_1$.
This is consistent with the fact that
only BPS states contribute to the index $I$.
When $\Delta_\Phi=1/2$ (\ref{fchiral}) agrees with the
corresponding function in \cite{Kim:2009wb}.

The letter index of vector multiplets is also obtained in a similar way.
Because any vector multiplet carries no flavor charges,
it is a function of only $s$, $e^{ia}$ and $x_2$.
The explicit form of $f_{\rm vector}$ is given by\cite{Kim:2009wb} 
\begin{equation}
f_{\rm vector}(e^{ia},x_2)=\sum_{\alpha\in G}
\left(-e^{i\alpha(a)}x_2^{2|\alpha(s)|}\right),
\label{fvector}
\end{equation}
where $\sum_{\alpha\in G}$ represents the summation
over all roots.

We also need to evaluate the
first factor in (\ref{pplambda}).
It is a monomial of the variables $e^{ia}$, $x_2$, and $z_i$.
We define $b_0$, $\epsilon_0$, and $q_{0i}$ by
\begin{eqnarray}
\exp\left(-\sum(-)^F\frac{z}{2}\right)=e^{ib_0(a)}x_2^{\epsilon_0}z_i^{q_{0i}}.
\end{eqnarray}
$\epsilon_0$ and $q_{0i}$ are zero-point contributions to the energy and
the flavor charges.
$b_0(a)$ is a linear function of $a$ which represents the
zero-point gauge charge.
We can derive $\epsilon_0$ from the total letter index
$f_{\rm tot}=f_{\rm chiral}+f_{\rm vector}$ by\cite{Kim:2009wb}
\begin{equation}
\epsilon_0=\frac{1}{2}\left.\frac{\partial f_{\rm tot}}{\partial x_2}\right|_{e^{ia}=x_2=z_i=1}
=\sum_\Phi (1-\Delta_\Phi)\sum_{\rho\in R_\Phi}|\rho(s)|
-\sum_{\alpha\in G}|\alpha(s)|.
\end{equation}
The zero-point flavor charges $q_{0i}$ are also obtained in the same way.
The result, however, diverges when we take the limit $x_2\rightarrow 1$.
\begin{equation}
q_{0i}=\frac{1}{2}\left.\frac{\partial f_{\rm tot}}{\partial z_i}\right|_{e^{ia}=z_i=1}
=-\sum_\Phi\sum_{\rho\in R_\Phi}F_i\left[\frac{1}{2(x_2-1)}+\left(\frac{1}{4}+|\rho(s)|\right)+{\cal O}(x_2-1)\right].
\label{q0iformal}
\end{equation}
We need some regularization.
Because (\ref{q0iformal}) does not depend on $\Delta_\Phi$,
it is plausible that
after an appropriate regularization
$q_{0i}$ does not depend on $\Delta_\Phi$.
Thus we take
zero-point charges for canonical fields,
\begin{equation}
q_{0i}=-\sum_\Phi\sum_{\rho\in R_\Phi}|\rho(s)|F_i.
\label{q0i}
\end{equation}
Similarly, $b_0(a)$ is given by
\begin{equation}
b_0(a)=-\sum_\Phi\sum_{\rho\in R_\Phi}|\rho(s)|\rho(a)
\label{b0}.
\end{equation}
(\ref{b0}) can be regarded as the $1$-loop correction to Chern-Simons terms.
This vanishes
when the matter representation is vector-like.

By collecting all components,
we obtain the following general formula for the index.
\begin{equation}
I(x_2,z_i)=\sum_sy_a^{2s_a}\int da e^{-S_{\rm CS}^{(0)}}e^{ib_0(a)}x_2^{\epsilon_0}z_i^{q_{0i}}
\exp\left[\sum_{m=1}^\infty\frac{1}{m}
f_{\rm tot}(e^{ima},x_2^m,z_i^m)
\right].
\end{equation}

Before ending this section
we comment on the integration measure associated with the Wilson line $a$.
The Wilson line $a$ is gauge fixed so that
$a$ is an element of Cartan subalgebra.
The associated Jacobian factor is
the Vandermonde determinant
\begin{equation}
J=\prod_{\alpha\in G,\alpha(s)=0}2i\sin\frac{\alpha(a)}{2}.
\label{jacc}
\end{equation}
Because the gauge group is broken by the magnetic flux $s$,
the product in (\ref{jacc})
is taken over only roots for unbroken gauge group,
which satisfy $\alpha(s)=0$.
This Jacobian factor can be rewritten in the form
\begin{equation}
J=\exp\left(\sum_{m=1}^\infty\frac{1}{m}f'(e^{ima})\right),
\end{equation}
with
\begin{equation}
f'(e^{ia})=\sum_{\alpha\in G,\alpha(s)=0}(-e^{-i\alpha(a)}).
\end{equation}
We included this contribution in the definition of
$f_{\rm vector}$ in (\ref{fvector}).
Although this simplifies the formula,
the expression (\ref{jacc}) is more useful
for actual computation.
In addition to the Jacobian factor
(\ref{jacc}) arising from the fixing of continuous
gauge symmetries,
we need a statistical factor associated with the Weyl group
of unbroken gauge symmetry.
For example, if the gauge group is $U(N)$ and
it is broken to $\prod U(N_k)$ by the flux $s$,
we should include the factor
$(\prod N_k!)^{-1}$ in the definition of the
integration measure $\int da$.

\section{Examples}\label{examples}
An ${\cal N}=2$ supersymmetric $U(1)$ gauge theory with
$N_f$ flavors $(Q_i,\wt Q_i)$ ($i=1,\ldots,N_f$) without superpotential
is known to be mirror to
a $U(1)^{N_f-1}$ quiver gauge theory\cite{Intriligator:1996ex,Aharony:1997bx,deBoer:1997ka}.
We refer to the QED with $N_f$ flavors by QED$_{N_f}$ and
its mirror by $\wt{\rm QED}_{N_f}$ for short.
In this section we compute the index for QED$_{N_f}$ and $\wt{\rm QED}_{N_f}$
with $N_f=1,2,3$,
and confirm the agreement of the indices in each mirror pair.

Let $V_i$ ($i=1,\ldots,N_f-1$) be the vector multiplets in $\wt{\rm QED}_{N_f}$.
The matter contents of $\wt{\rm QED}_{N_f}$ are
$N_f$ singlets $S_i$ ($i=1,\ldots,N_f$) and
$N_f$ pairs of bi-fundamental chiral multiplets $(q_i,\wt q_i)$ ($i=1,\ldots,N_f$).
$q_i$ and $\wt q_i$ couple to $V_i-V_{i-1}$ with opposite charges,
where $V_0=V_{N_f}=0$ is understood.
The superpotential is
\begin{equation}
W_{\wt{\rm QED}_{N_f}}=\sum_{i=1}^{N_f}\wt q_iq_iS_i.
\end{equation}

Let us first consider QED$_{N_f}$.
This theory has $U(N_f)\times U(N_f)$ symmetry rotating $Q_i$ and $\wt Q_i$
separately.
We introduce the Cartan generators $F_i$ and $F_i'$
so that for each $i$
they act only on the $i$-th flavor $(Q_i,\wt Q_i)$.
Namely, we adopt the charge assignments
\begin{equation}
F_i(Q_j)=F_i(\wt Q_j)=\delta_{ij},\quad
F'_i(Q_j)=-F'_i(\wt Q_j)=\delta_{ij}.
\end{equation}
Corresponding to these charges,
we introduce variables $t_i$ and $t_i'$,
and insert the operator $t_i^{F_i}t_i'^{F_i'}$ in the
definition of the index.
Because the rotation by $F_1'+\cdots+F_{N_f}'$ is the same as the $U(1)$ gauge
transformation,
the set of $N_f$ variables $t_i'$ is redundant,
and the index is invariant under
the rescaling $t_i'\rightarrow at_i'$.
We set $t'_{N_f}=1$ by using this rescaling.

Let us define the function
\begin{equation}
f_\Delta(s,x,t)=\frac{tx^{2|s|+\Delta}-t^{-1}x^{2|s|+2-\Delta}}{1-x^2}.
\end{equation}
(We here use $x$ instead of $x_2$.)
The letter index of QED$_{N_f}$ is
\begin{equation}
f_{{\rm QED}_{N_f}}(s,e^{ia},x,t_i,t_i')=\sum_{i=1}^{N_f}
[f_h(s,x,e^{ia}t_it'_i)
+f_h(s,x,e^{-ia}t_it_i'^{-1})],
\end{equation}
where $h$ is the Weyl weight of $Q_i$ and $\wt Q_i$;
$h=\Delta(Q_i)=\Delta(\wt Q_i)$.
With this letter index, the index of QED$_{N_f}$ is given by
\begin{align}
I_{{\rm QED}_{N_f}}(x,t_i,t_i',t'')=&\sum_{s\in{\bf Z}/2}t''^{2s}\int\frac{da}{2\pi}
x^{2N_f(1-h)|s|}(t_1\cdots t_{N_f})^{-2|s|}
\nonumber\\
&\hspace{2em}
\exp\left(\sum_{m=1}^\infty\frac{1}{m}
f_{{\rm QED}_{N_f}}(s,e^{ima},x^m,t_i^m,t_i'^m)\right),
\end{align}
where we introduce the variable $t''$ for the monopole charge $2s$.
We define the operator $F''$ counting the monopole charge for later use.

Although we focus only on the Cartan part in the computation of the index,
QED$_{N_f}$ actually has $SU(N_f)^2\times U(1)^2$ global symmetry,
and thus the index must be expanded by $SU(N_f)$ characters.
We need some variable changes for this character expansion.
We separate diagonal part from $t_i$ and define $t_0$ and $\wt t_i$ by
\begin{equation}
t_0=(t_1t_2\cdots t_{N_f})^{1/{N_f}},\quad
\wt t_i=\frac{t_i}{t_{N_f}}\quad(i=1,\ldots,N_f-1).
\label{u1part}
\end{equation}
We rearrange the two sets of $N_f-1$ variables $\wt t_i$ and
$t_i'$ by
\begin{equation}
u_i=\wt t_it_t',\quad
u'_i=\frac{\wt t_i}{t_t'}.
\label{susu}
\end{equation}
The two sets of variables $u_i$ and $u_i'$ correspond to
the two $SU(N_f)$ global symmetries.
After this change of variables, the index is expanded by
$SU(N_f)\times SU(N_f)$ characters $\chi_R(u_i)\chi_{R'}(u_i')$,
where $R$ and $R'$ are $SU(N_f)$ representations.

Next, let us consider $\wt{\rm QED}_{N_f}$.
This theory also has $2N_f$ global $U(1)$ symmetries as QED$_{N_f}$:
$F_i$ ($i=1,\ldots,N_f$), $F_i'$ ($i=1,\ldots,N_f-1$), and $F''$.
The symmetries generated by $F_i$ are realized as symmetries
acting on elementary fields just as in QED$_{N_f}$.
The charge assignments are
\begin{equation}
F_i(q_j)=F_i(\wt q_j)=-\delta_{ij},\quad
F_i(S_j)=2\delta_{ij}.
\end{equation}
The roles of $F_i'$ and $F''$ in $\wt{\rm QED}_{N_f}$ are interchanged 
compared to those in ${\rm QED}_{N_f}$:
in $\wt{\rm QED}_{N_f}$,
$F_i'$ are monopole charges and $F''$ is a perturbative symmetry.
The charge assignments for $F''$ are
\begin{equation}
F''(q_j)=\delta_{jI},\quad
F''(\wt q_j)=-\delta_{jI},\quad
F''(S_j)=0.\label{fppdef}
\end{equation}
To define $F''$ we need to fix $I(=1,\ldots,N_f)$ specifying
fields on which $F''$ non-trivially acts.
Two $F''$ with different choices of $I$ are the same up to gauge transformation.
In the following we use $I=1$.

It is known that $S_i$ in $\wt{\rm QED}_{N_f}$ corresponds to
the composite field $\wt Q_iQ_i$ for each $i$.
With this relation, we obtain the Weyl weight of fields in $\wt{\rm QED}_{N_f}$
\begin{equation}
\Delta(q_i)=\Delta(\wt q_i)=1-h,\quad
\Delta(S_i)=2h.
\end{equation}
The letter index of $\wt{\rm QED}_{N_f}$ is
\begin{align}
&f_{\wt{\rm QED}_{N_f}}(s_i,e^{ia_i},x,t_i,t'')
\nonumber\\
&=\sum_{i=1}^{N_f}\Big[
f_{1-h}(s_i-s_{i-1},x,e^{i(a_i-a_{i-1})}t_i^{-1}t''_i)
\nonumber\\&
\hspace{4em}
+f_{1-h}(s_i-s_{i-1},x,e^{-i(a_i-a_{i-1})}t_i^{-1}t''^{-1}_i)
+f_{2h}(0,x,t_i^2)
\Big],
\label{eq76}
\end{align}
where $s_0=s_{N_f}=a_0=a_{N_f}=0$ is understood.
Corresponding to (\ref{fppdef}) with $I=1$, $t''_i$
in (\ref{eq76}) are defined by
\begin{equation}
t''_1=t'',\quad
t''_i=1\quad\mbox{for $i\geq2$}.
\end{equation}
The index is given by
\begin{align}
&I_{\wt{\rm QED}_{N_f}}(x,t_i,t_i',t'')
\nonumber\\
&=
\prod_{i=1}^{N_f-1}\left(\sum_{s_i\in{\bf Z}/2}
\left(\frac{t_i'}{t_{i+1}'}\right)^{2s_i}
\int\frac{da_i}{2\pi}\right)
\prod_{i=1}^{N_f}\left(x^{2h|s_i-s_{i-1}|}t_i^{2|s_i-s_{i-1}|}\right)
\nonumber\\&
\hspace{4em}
\exp\left(
\sum_{m=1}^\infty\frac{1}{m}
f_{\wt{\rm QED}_{N_f}}(s_i,e^{ima_i},x^m,t_i^m,t''^m)
\right),
\end{align}
where we introduce $N_f$ variables $t_i'$ for $N_f-1$ monopole charges
so that there is the same redundancy as in the QED$_{N_f}$.
This index is invariant under the rescaling $t_i'\rightarrow at_i'$,
and we fix this degrees of freedom by $t_{N_f}'=1$.

Now we have two indices which should agree to each other.
Let us confirm this agreement for $N_f=1,2,3$ by numerical
computation.

\subsection{$N_f=1$}

Let us first consider the $N_f=1$ case.
In this case, the mirror theory $\wt{\rm QED}_1$ does not
have gauge symmetry, and is
the Wess-Zumino model with three chiral multiplets
$q$, $\wt q$, $S$ interacting through the superpotential $W=\wt qSq$.
In this special case, thanks to the
$S_3$ permutation symmetry among $q$, $\wt q$, and $S$,
we can uniquely determine the Weyl weight of these fields as $\Delta=2/3$.
This corresponds to $h=1/3$.
However, we will not use this value and keep $h$ unfixed.
As we will see below, the two indices agree regardless of
the value of $h$.

The letter index for QED$_1$ is
\begin{equation}
f_{{\rm QED}_1}(s,x,e^{ia},t_1)=f_h(s,x,e^{ia}t_1)+f_h(s,x,e^{-ia}t_1).
\end{equation}
The complete index is
\begin{align}
&I_{{\rm QED}_1}(x,t_1,t'')\nonumber\\
&=\sum_{s\in{\bf Z}/2}t''^{2s}\int\frac{da}{2\pi}x^{2(1-h)|s|}t_1^{-2|s|}
\exp\left(\sum_{n=1}^\infty\frac{1}{n}f_{{\rm QED}_1}(s,x^n,e^{ina},t_1^n)\right).
\end{align}

On the other hand, the index of the mirror theory, $\wt{\rm QED}_1$, is given by
\begin{equation}
I_{\wt{\rm QED}_1}(x,t_1,t'')=
\exp\left(\sum_{n=1}^\infty\frac{1}{n}f_{\wt{\rm QED}_1}(x^n,t_1^n,t''^n)\right),
\end{equation}
where $f_{\wt{\rm QED}_1}(x,t_1,t'')$ is the letter index
\begin{equation}
f_{\wt{\rm QED}_1}(x,t_1,t'')
=f_{1-h}(0,x,t''t_1^{-1})
+f_{1-h}(0,x,t''^{-1}t_1^{-1})
+f_{2h}(0,x,t_1^2).
\end{equation}

Although we cannot prove the equality of these two
indices analytically,
it is easy to show that the series expansions of them agree
up to some order.
In the computation without fixing $h$,
it is convenient to define
\begin{equation}
y=x^h,\quad
z=x^{1-h}.
\end{equation}
We can easily find that both indices contain
only terms with non-negative power of $y$ and $z$.
We have confirmed that they agree up to
${\cal O}(y^{16})$ and ${\cal O}(z^{16})$.
Due to limitations of space we only show the first several terms of the series expansion.
\begin{align}
&
I_{{\rm QED}_1}(x,t_1,t'')=I_{\wt{\rm QED}_1}(x,t_1,t'')\nonumber\\
&=\bigg(1
+\left(\frac{1}{t''t_1}+\frac{t''}{t_1}\right)z
+\left(\frac{1}{t''^2t_1^2}+\frac{t''^2}{t_1^2}\right)z^2
+\left(\frac{1}{t''^3t_1^3}+\frac{t''^3}{t_1^3}\right)z^3
\nonumber\\&\hspace{17em}
+\left(\frac{1}{t''^4t_1^4}+\frac{t''^4}{t_1^4}\right)z^4
+{\cal O}(z^5)\bigg)\nonumber\\
&+\left(t_1^2-2z^2+\frac{z^4}{t_1^2}+{\cal O}(z^5)\right)y^2\nonumber\\
&+\left(t_1^4+\left(\frac{t_1}{t''}+t''t_1\right)z^3-3z^4+{\cal O}(z^5)\right)y^4\nonumber\\
&+{\cal O}(y^5).
\end{align}
We also give the index with $t_1=t''=1$ substituted up to ${\cal O}(y^{16})$ and ${\cal O}(z^{16})$ terms:
\begin{align}
&I_{{\rm QED}_1}(x,t_1=t''=1)=I_{\wt{\rm QED}_1}(x,t_1=t''=1)\nonumber\\
&=
(1+2z+2z^2+2z^3+2z^4+2z^5+2z^6+2z^7+2z^8+2z^9+2z^{10}
\nonumber\\&\hspace{5em}
+2z^{11}+2z^{12}+2z^{13}+2z^{14}+2z^{15}+{\cal O}(z^{16}))\nonumber\\
&+(1-2z^2+z^4+{\cal O}(z^{16}))y^2\nonumber\\
&+(1+2z^3-3z^4-4z^5+2z^6+2z^7+{\cal O}(z^{16}))y^4\nonumber\\
&+(1-2z^4+4z^5+2z^6-8z^7-3z^8+4z^9+2z^{10}+{\cal O}(z^{16}))y^6\nonumber\\
&+(1+2z^5-4z^6+2z^7+8z^8-8z^9-10z^{10}+2z^{11}+5z^{12}+2z^{13}+{\cal O}(z^{16}))y^8\nonumber\\
&+(1-2z^6+4z^7-5z^8-4z^9+18z^{10}-18z^{12}-4z^{13}+4z^{14}+4z^{15}+{\cal O}(z^{16}))y^{10}\nonumber\\
&+(1+2z^7-4z^8+6z^9-16z^{11}+18z^{12}+14z^{13}-24z^{14}-12z^{15}+{\cal O}(z^{16}))y^{12}\nonumber\\
&+(1-2z^8+4z^9-6z^{10}+4z^{11}+8z^{12}-28z^{13}+14z^{14}+40z^{15}+{\cal O}(z^{16})y^{14}\nonumber\\
&+{\cal O}(y^{16}).
\end{align}

\subsection{$N_f=2$}
In the case of $N_f=2$,
the letter index of the QED$_2$ is
\begin{align}
f_{{\rm QED}_2}(s,e^{ia},x,t_1,t_2,t_1')
=&f_h(s,x,e^{ia}t_1t_1')
+f_h(s,x,e^{-ia}t_1t_1'^{-1})
\nonumber\\&
+f_h(s,x,e^{ia}t_2)
+f_h(s,x,e^{-ia}t_2),
\end{align}
and the complete index is
\begin{align}
I_{{\rm QED}_2}(x,t_1,t_2,t_1',t'')&=\sum_{s\in{\bf Z}/2}t''^{2s}\int\frac{da}{2\pi}
x^{4(1-h)|s|}t_1^{-2|s|}t_2^{-2|s|}
\nonumber\\&\hspace{1em}
\exp\left(\sum_{m=1}^\infty\frac{1}{m}
f_{{\rm QED}_2}(s,e^{ima},x^m,t_1^m,t_2^m,t_1'^m)\right).
\end{align}
For the mirror theory $\wt{\rm QED}_2$,
the letter index is
\begin{align}
&f_{\wt{\rm QED}_2}(s,e^{ia},x,t_1,t_2,t'')
\nonumber\\
&=f_{1-h}(s,x,e^{ia}t_1^{-1}t'')
+f_{1-h}(s,x,e^{-ia}t_1^{-1}t''^{-1})
+f_{2h}(0,x,t_1^2)
\nonumber\\&
+f_{1-h}(s,x,e^{-ia}t_2^{-1})
+f_{1-h}(s,x,e^{ia}t_2^{-1})
+f_{2h}(0,x,t_2^2),
\end{align}
and the complete index is
\begin{align}
I_{\wt{\rm QED}_2}(x,t_1,t_2,t'_1,t'')
=&\sum_{s\in{\bf Z}/2}t_1'^{2s}\int\frac{da}{2\pi}
x^{4h|s|}t_1^{2|s|}t_2^{2|s|}
\nonumber\\&
\exp\left(\sum_{m=1}^\infty\frac{1}{m}
f_{\wt{\rm QED}_2}(s,e^{ima},x^m,t_1^m,t_2^m,t''^m)
\right).
\end{align}
We have confirmed that the two indices agree to each other up to ${\cal O}(y^{14})$ and ${\cal O}(z^{14})$ terms.
A part of the series expansion is
\begin{align}
&I_{{\rm QED}_2}(x,t_1,t_2,t_1',t'')
=I_{\wt{\rm QED}_2}(x,t_1,t_2,t_1',t'')\nonumber\\
&=\left(1+\frac{1}{t_1t_2}\left(t''+\frac{1}{t''}\right)z^2
+{\cal O}(z^4)\right)\nonumber\\
&+\left(\left(t_1^2+t_2^2+t_1t_2\left(t_1'+\frac{1}{t_1'}\right)\right)
+\left(-4-\left(\frac{t_1}{t_2}+\frac{t_2}{t_1}\right)\left(t_1'+\frac{1}{t_1'}\right)\right)z^2
+{\cal O}(z^4)\right)y^2\nonumber\\
&+{\cal O}(y^4).
\end{align}
After the variable changes (\ref{u1part}) and (\ref{susu}),
this is expanded by the $SU(2)$ character
\begin{equation}
\chi_s(u)=u^{-s}+\cdots+u^s
\end{equation}
as
\begin{align}
&I_{{\rm QED}_2}(x,t_0,t'',u_1,u_1')
=I_{\wt{\rm QED}_2}(x,t_0,t'',u_1,u_1')\nonumber\\
&=1
+\frac{z^2}{t_0^2}\left(t''+\frac{1}{t''}\right)
+\frac{z^4}{t_0^4}\left(t''^2+\frac{1}{t''^2}\right)
+{\cal O}(z^6)
\nonumber\\&
+\left(\chi_{\frac{1}{2}}(u_1)\chi_{\frac{1}{2}}(u_1')
-(\chi_1(u_1)+\chi_1(u_1')+2)\frac{z^2}{t_0^2}
+\chi_{\frac{1}{2}}(u_1)\chi_{\frac{1}{2}}(u_1')\frac{z^4}{t_0^4}+{\cal O}(z^6)\right)t_0^2y^2
\nonumber\\&
+\bigg(\chi_1(u_1)\chi_1(u_1')
-\left(
\chi_{\frac{3}{2}}(u_1)\chi_{\frac{1}{2}}(u_1')
\chi_{\frac{1}{2}}(u_1)\chi_{\frac{3}{2}}(u_1')\right)\frac{z^2}{t_0^2}
\nonumber\\&\hspace{2em}
+\left(\left(t''+\frac{1}{t''}\right)\chi_{\frac{1}{2}}(u_1)\chi_{\frac{1}{2}}(u_1')
+\chi_1(u_1)+\chi_1(u_1')-3\right)\frac{z^4}{t_0^4}+{\cal O}(z^6)\bigg)t_0^4y^4
\nonumber\\&
+{\cal O}(y^6).
\end{align}
We also give the index with $t_i=t_1'=t''=1$ substituted.
\begin{eqnarray}
&&I_{{\rm QED}_2}(x,t_i=t_1'=t''=1)
=I_{\wt{\rm QED}_2}(x,t_i=t_1'=t''=1)\nonumber\\
&&=( 1+ 2z^2+ 2z^4+ 2z^6+  2z^8+  2z^{10}+  2z^{12}+{\cal O}(z^{14}))\nonumber\\
&&+( 4- 8z^2+ 4z^4+{\cal O}(z^{14})) y^2\nonumber\\
&&+( 9-16z^2+14z^4-16z^6+  9z^8+{\cal O}(z^{14})) y^4\nonumber\\
&&+(16-24z^2-16z^4+64z^6- 56z^8+  8z^{10}+  8z^{12}+{\cal O}(z^{14})) y^6\nonumber\\
&&+(25-32z^2-32z^4+66z^6-  3z^8- 36z^{10}+  2z^{12}+{\cal O}(z^{14})) y^8\nonumber\\
&&+(36-40z^2-48z^4+16z^6+160z^8-216z^{10}+160z^{12}+{\cal O}(z^{14})) y^{10}\nonumber\\
&&+(49-48z^2-64z^4+118z^8+164z^{10}-542z^{12}+{\cal O}(z^{14})) y^{12}\nonumber\\
&&+{\cal O}(y^{14}).
\end{eqnarray}

\subsection{$N_f=3$}
As the last example, let us compute the indices for $N_f=3$.
We have computed the indices for QED$_3$ and $\wt{\rm QED}_3$
and confirmed that they agree to each other up to ${\cal O}(y^{13})$
and ${\cal O}(z^{13})$ terms.
We show only a part of the series expansion.
\begin{align}
&I_{{\rm QED}_3}(x,t_i,t_i',t'')=I_{\wt{\rm QED}_3}(x,t_i,t_i',t'')\nonumber\\
&=1+\left(t_1^2+t_2^2+t_3^2+t_1t_3\left(t_1'+\frac{1}{t_1'}\right)
+t_2t_3\left(t_2'+\frac{1}{t_2'}\right)
+t_1t_2\left(\frac{t_2'}{t_1'}+\frac{t_1'}{t_2'}\right)\right)y^2
\nonumber\\&
-\bigg(6
+\left(t_1'+\frac{1}{t_1'}\right)\left(\frac{t_1}{t_3}+\frac{t_3}{t_1}\right)
+\left(t_2'+\frac{1}{t_2'}\right)\left(\frac{t_2}{t_3}+\frac{t_3}{t_2}\right)
\nonumber\\&\hspace{18em}
+\left(\frac{t_2'}{t_1'}+\frac{t_1'}{t_2'}\right)\left(\frac{t_1}{t_2}+\frac{t_2}{t_1}\right)
\bigg)y^2z^2
\nonumber\\&
+\frac{1}{t_1t_2t_3}\left(t''+\frac{1}{t''}\right)z^3
+{\cal O}(y^4)+{\cal O}(z^4).
\end{align}
After the variable changes (\ref{u1part}) and (\ref{susu}),
this can be rewritten as
\begin{align}
&I_{{\rm QED}_3}(x,t_0,t'',u_i,u'_i)=I_{\wt{\rm QED}_3}(x,t_0,t'',u_i,u'_i)
\nonumber\\
&=1
+\frac{1}{t_0^3}\left(t''+\frac{1}{t''}\right)z^3
+t_0^2\chi_{(1,0)}(u_i)\chi_{(1,0)}(u'_i)y^2
\nonumber\\&
-(\chi_{(1,1)}(u_i)+\chi_{(1,1)}(u'_i)+2)y^2z^2
+\frac{1}{t_0^2}\chi_{(0,1)}(u_i)\chi_{(0,1)}(u_i')y^2z^4
\nonumber\\&
+t_0^4\chi_{(2,0)}(u_i)\chi_{(2,0)}(u_i')y^4
-t_0^2(\chi_{(2,1)}(u_i)\chi_{(1,0)}(u_i')+\chi_{(1,0)}(u_i)\chi_{(2,1)}(u_i'))y^4z^2
\nonumber\\&
+t_0^2(\chi_{(3,0)}(u_i)+\chi_{(3,0)}(u_i')+\chi_{(1,1)}(u_i)\chi_{(1,1)}(u_i')-3)y^4z^4
\nonumber\\&
+\frac{1}{t_0}\left(t''+\frac{1}{t''}\right)\chi_{(1,0)}(u_i)\chi_{(1,0)}(u'_i)y^4z^5
+{\cal O}(y^6)+{\cal O}(z^6),
\end{align}
where the $SU(3)$ character $\chi_{(m,n)}$ is defined so that
it is given for the fundamental and the anti-fundamental representations as
\begin{equation}
\chi_{(1,0)}(u_i)=\frac{1}{(u_1u_2)^{1/3}}(1+u_1+u_2),\quad
\chi_{(0,1)}(u_i)=(u_1u_2)^{1/3}\left(1+\frac{1}{u_1}+\frac{1}{u_2}\right).
\end{equation}
The index with $t_i=t_i'=t''=1$ substituted is
\begin{eqnarray}
&&I_{{\rm QED}_3}(x,t_i=t_i'=t''=1)=I_{\wt{\rm QED}_3}(x,t_i=t_i'=t''=1)
\nonumber\\
&&=(1+2 z^3+2 z^6+2z^9+2z^{12}+{\cal O}(z^{13}))\nonumber\\
&&+y^2(9-18z^2+9z^4+{\cal O}(z^{13}))\nonumber\\
&&+y^4(36-90z^2+81z^4+18z^5-36z^6-36z^7+9z^8+18z^9+{\cal O}(z^{13}))\nonumber\\
&&+y^6(100-252z^2+153z^4+88z^6+36z^7-126z^8-72z^9+54z^{10}
\nonumber\\&&\hspace{5cm}
+36z^{11}-35z^{12}+{\cal O}(z^{13}))\nonumber\\
&&+y^8(225-540z^2+153z^4+612z^6+72z^7-684z^8-162z^9+288z^{10}
\nonumber\\&&\hspace{5cm}
+126z^{11}-36z^{12}+{\cal O}(z^{13}))\nonumber\\
&&+y^{10}(441-990z^2+9z^4+1368z^6-612z^8+216z^9-918z^{10}-648z^{11}
\nonumber\\&&\hspace{5cm}
+1107z^{12}+{\cal O}(z^{13}))\nonumber\\
&&+y^{12}(784-1638z^2-351z^4+2304z^6+288z^8+200z^9-2754z^{10}
\nonumber\\&&\hspace{5cm}
-198z^{11}+1140z^{12}+{\cal O}(z^{13}))\nonumber\\
&&+{\cal O}(y^{13}).
\end{eqnarray}

\section{Discussions}\label{discussion}
We derived a general formula for
an index 
for three-dimensional ${\cal N}=2$ superconformal field theories
with general R-charge assignments
by using the localization procedure in ${\bf S}^2\times{\bf S}^1$.
As an application, we computed the index for theories in
a few mirror pairs, and confirmed that the
indices agree in each mirror pairs.

Although the gauge groups
in the examples we discussed in this paper are all Abelian,
the formula we derived can apply to theories with
arbitrary gauge groups.
In particular, it is possible to compute the index
in the large $N$ limit.
It enable us to study AdS$_4$/CFT$_3$ correspondence
for a large class of ${\cal N}=2$ quiver gauge theories.
Such an analysis is performed in
\cite{Bhattacharya:2008zy,Bhattacharya:2008bja,Kim:2009wb}
for the ABJM model and the complete agreement
of the gauge theory index and the gravity index
is confirmed.
Similar analysis is also done for ${\cal N}=3,4,5$ Chern-Simons theories
in \cite{Choi:2008za,Imamura:2009hc,Kim:2010vwa}.
It would be interesting to extend such an analysis to
more general ${\cal N}=2$ theories including chiral multiplets
with non-canonical R-charges.

At least in the examples we consider in this paper,
the index does not give additional information
for the R-charge assignment than what is obtained from
simple operator matchings.
This is natural because the index contains information
about only the BPS sector.
It would be necessary to use the $Z$-extremization recently proposed
in \cite{Jafferis:2010un} to obtain the R-charge assignment
in the infrared fixed point.

An advantage of the index over the partition function is
that in the index we can separate the contribution
of monopole operators.
In AdS$_4$/CFT$_3$ correspondence, monopole operators
play important roles.
On the gravity side, a part of monopole operators are expected to
correspond to M2-branes wrapped on two-cycles\cite{Imamura:2008ji}.
The relation between monopole charges and M2-brane wrapping
numbers are highly non-trivial.
To address such an issue, 
the index is useful to establish the one-to-one map
between monopole charges in a CFT and corresponding quantum numbers on
the gravity side\cite{Imamura:2010sa}.

\section*{Acknowledgements}
Y.I. was supported in part by
Grant-in-Aid for Young Scientists (B) (\#19740122) from the Japan
Ministry of Education, Culture, Sports,
Science and Technology.
S.Y. was supported by
the Global COE Program ``the Physical Sciences Frontier'', MEXT, Japan.

\paragraph{Note added:}
When we revised this paper in March 2011
we improved the analysis in Section 5 by introducing
chemical potentials for $U(1)$ global symmetries.
After completion of the revision, a paper appeared\cite{Krattenthaler:2011da}
which also studies the index with such chemical potentials for theories
studied in Section 5.

\end{document}